\pdfoutput=1
\documentclass[twocolumn]{aastex6}

\usepackage{amsmath}
\usepackage{ulem}
\usepackage{color}

\renewcommand{\textbf}{}

\shorttitle{MHD simulation of the formation of a long flux rope}
\shortauthors{He et al.}

\begin{document}

\title{Data-driven MHD Simulation of the Formation and Initiation of a
  Large-scale Pre-flare Magnetic Flux Rope in Solar Active Region
  12371}

\author{Wen He\altaffilmark{1,3},
  Chaowei Jiang*\altaffilmark{1}
  Peng Zou\altaffilmark{1},
  Aiying Duan\altaffilmark{2},
  Xueshang Feng\altaffilmark{1},
  Pingbing Zuo\altaffilmark{1},
  Yi Wang\altaffilmark{1}}

\altaffiltext{1}{Institute of Space Science and Applied Technology,
  Harbin Institute of Technology, Shenzhen 518055, China;
  \href{mailto:chaowei@hit.edu.cn}{chaowei@hit.edu.cn}}

\altaffiltext{2}{School of Atmospheric Sciences, Sun Yat-sen
  University, Zhuhai, Guangdong 519082, China}

\altaffiltext{3}{Center for Space Plasma and Aeronomic Research, The
  University of Alabama in Huntsville, Huntsville, AL 35899, USA}

\begin{abstract}
  Solar eruptions are the most powerful drivers of space weather. To
  understand their cause and nature, it is crucial to know how the
  coronal magnetic field evolves before eruption. Here we study the
  formation process of a relatively large-scale magnetic flux rope
  (MFR) in active region NOAA~12371 that erupts with a major flare and
  coronal mass ejection on 2015 June 21. A data-driven numerical
  magnetohydrodynamic model is employed to simulate three-dimensional
  coronal magnetic field evolution of one-day duration before the
  eruption. Comparison between the observed features and our modeled
  magnetic field discloses how the pre-eruption MFR forms. Initially,
  the magnetic field lines were weakly twisted as being simple sheared
  arcades. Then a long MFR was formed along the polarity inversion
  line due to the complex photospheric motion, which is mainly
  shearing rather than twisting. The presence of the MFR is evidenced
  by a coherent set of magnetic field lines with twist number above
  unity. Below the MFR a current sheet is shown in the model,
  suggesting that tether-cutting reconnection plays a key role in the
  MFR formation. The MFR's flux grows as more and more field lines are
  twisted due to continuous injection of magnetic helicity by the
  photospheric motions. Meanwhile, the height of the MFR's axis
  increases monotonely from its formation. By an analysis of the decay
  index of its overlying field, we suggest that it is because the MFR
  runs into the torus instability regime and becomes unstable that
  finally triggers the eruption.
\end{abstract}

\keywords{Sun: corona; Sun: magnetic fields; Sun: coronal mass
  ejections (CMEs); Sun: flares; magnetohydrodynamics (MHD); methods:
  numerical}

\section{Introduction}
\label{sec:intro}

Large-scale eruptions occurring in the solar atmosphere can release a
vast amount of energy up to 10$^{32}$~erg in \textbf{tens of} minutes
and may severely affect the space environment around the Earth. Such
phenomena including flares, filament eruptions and coronal mass
ejections (CMEs) are driven commonly by the Sun's magnetic field
evolution. In particular, the magnetic field plays a dominant role in
the solar corona because the plasma $\beta$, i.e., ratio of gas
pressure to magnetic pressure, is often very small. The coronal
magnetic field can be strongly stressed by photospheric flux
emergences and motions, and excess or free magnetic energy is
accumulated until a catastrophic release occurs, which powers solar
eruption events~\citep{2004Aschwanden}. During solar eruptions,
magnetic reconnection is thought to be the key mechanism that converts
magnetic free energy to radiation and energetic particle
acceleration~\citep{2002Priest}. Meanwhile, it cuts parts of the
connection of the magnetic flux with the Sun and allows \textbf{a huge
  amount of magnetized plasma to be ejected} into interplanetary space
as coronal mass ejections.
Since variation of magnetic field topology has a
close relationship with magnetic reconnection, it is essential to
understand the evolution of magnetic configuration in the corona to
figure out the nature and cause of solar eruptions.

Direct and accurate measurement of the magnetic field is less
accessible in the chromosphere and corona than in the photosphere due
to the low density and high temperature, which gives rise to many
theoretical models being proposed. For example, the standard CME/flare
model is frequently mentioned \citep{1964Carmichael, 1966Sturrock,
  1974Hirayama,1976KoppPneuman}. It introduces a conceptual scenario
that a magnetic flux rope (MFR) in the corona \textbf{is ejected} into
interplanetary space. An MFR is a bundle of twisted magnetic field
lines lying above the polarity inversion line (PIL) of photospheric
magnetic field, with two legs anchored at the photosphere and, some
\textbf{parts} of field lines of the MFR may be manifested as
different observable features such as \textbf{filaments, Sigmoids or
  hot channels}~\citep{2013ChengX,2014ChengX}. With the rising of the
MFR, its overlying field is strongly stretched and squeezed below the
MFR, where an electric current sheet \textbf{is} formed and
reconnection sets in. Then, part of the magnetic energy released in
the reconnection tracks the newly reconnected field lines to the
chromosphere and results in two parallel flare ribbons at the
footpoints of these field lines~\citep{2017Benz}.

However, these theoretical models are idealized or hypothetical
simplification of the realistic solar eruptions that is much more
complex and elusive than what the standard model shows
(e.g. \citealt{2018Jianga}). For example, the nature of the
pre-eruption configuration is still elusive. There are intensive
debates on whether MFR exists before flare or forms during flare
\citep{2000Forbes,2001Moore, 2011Chen}. A conclusive answer to this
question would provide a specific guidance to our understanding of
solar eruptions. Although a lot of \textbf{evidence is} found for that
MFR could exist prior to eruptions~\citep{2017ChengX}, there is still
no consensus on how and where an MFR can form. One supposition is that
the MFR can bodily emerge from below the photosphere by buoyancy
\citep{2001Fan, 2008Martinez, 2004Magara, 2009Archontis}. The other
supposes that the MFR can be built up directly in the corona via
magnetic reconnection prior to the eruption \citep{1989Ballegooijen,
  2010Aulanier, 2003aAmari}. \textbf{Moreover, it was claimed that
  some MFRs might be formed during eruption~\citep{Ouyang2017}}.

\textbf{Despite of the fact that routine observations of the
  photospheric magnetograms have been made in the}
past decades, reliable measurement of the full 3D magnetic field
\textbf{in the solar atmosphere} is still unavailable. Numerical
simulations based on the magnetohydrodynamics (MHD) model prove to be
a powerful tool to reproduce the time-dependent, nonlinear evolution
process of the 3D magnetic configuration and investigate the dynamic
evolution of solar eruptions. For instance, the formation of an MFR
directly in the corona and its eruption have been extensively studied
\citep{2010Aulanier,2012Aulanier}. However, such idealized
configuration of MFR might not be able to characterize the realistic
case in the solar corona. In order to account for the complexity and
evolution of the magnetic configuration in the real scene, a detailed
and accurate description of evolving magnetic field is
required. Realistic simulations of solar eruptions driven directly or
constrained by photospheric magnetograms provide an important way to
this end (e.g., \citealt{Wu2006, 2012Cheung, 2013Jianga}). Very
recently, such data-driven numerical simulations are becoming an more
and more active research field for solar eruptions. For instance,
\citet{2016Jiang} developed a data-driven MHD model that
self-consistently follows the time-sequence of
observations. \citet{2019Nayak} studied the magnetic reconnection
process of a blowout jet and a flare with a data-constrained MHD
simulation. \citet{2019cheung} presented a comprehensive radiative MHD
simulation of a solar flare to capture the process from emergence to
eruption. Time-dependent photospheric electric field and plasma flow
data \textbf{were} used by \citet{2019Hayashi} to conduct a
data-driven MHD simulation for solar active
region. \citet{2019Pomoell} analyzed the coronal response to the
driving electric fields as boundary conditions of data-driven
magnetofrictional simulation for the evolution of coronal magnetic
field. \citet{2019Guoyang} recently developed a data-driven MHD model
using the zero-$\beta$ approximation and successfully simulated an MFR
eruption in consistent with multi-wavelength observations.

In this paper, we conduct a data-driven MHD modelling study for the
formation process of an pre-eruption MFR, which helps to identify the
mechanisms of its build-up process and \textbf{initiation}. The
eruption event occurred in NOAA AR~12371 on 2015 June 21. Our
data-driven MHD simulation reproduces the dynamic evolution of the 3D
magnetic field covering one day before the flare onset time. The
simulation clearly demonstrates the creation of a large-scale
pre-flare MFR in the corona and its evolution until before the
eruption. By comparison with observations and previous
studies~\citep{2017Vemareddy, 2018Vemareddy}, we found that this MFR
is consistent with a long hot channel and filament observed by
SDO/AIA. We further performed a detailed analysis of the building-up
process, magnetic energy evolution and triggering mechanism of the
MFR. The remainder of this paper is organized as follows. Data and
method are presented in Section \ref{1}, then we analyze the evolution
of magnetic configuration in Section \ref{2} by observations and
simulation results, and we conclude in Section \ref{3}.

\begin{figure*}
  \centering
  \includegraphics[width=0.9\textwidth]{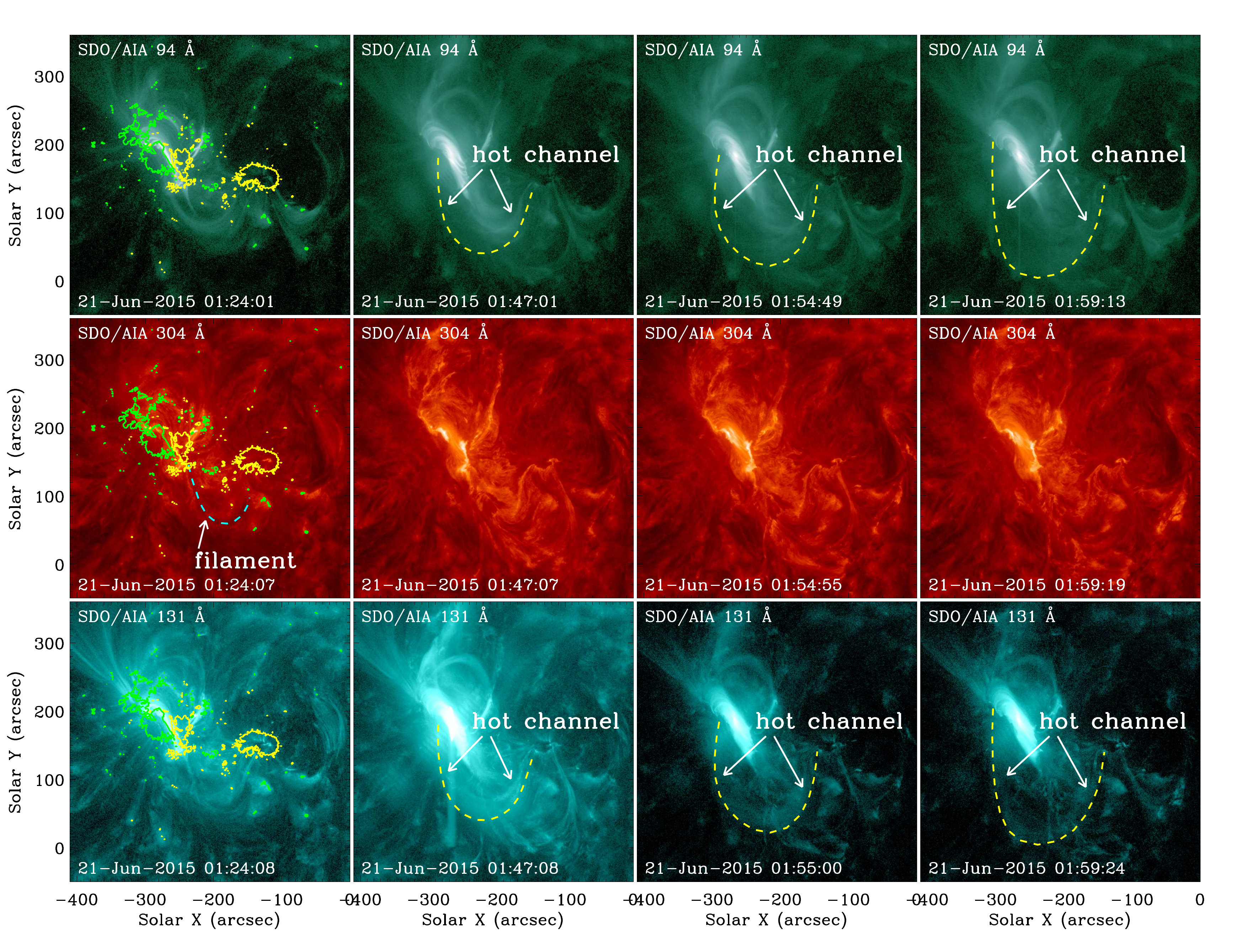}
  \caption{\textbf{SDO/AIA observations of evolution in AR 12371
      immediately before and during the eruption on early 2015 June
      21. Each panel has the same field of view. From top to bottom
      are respectively 94 \AA{}, 304 \AA{} and 131 \AA{}.  The hot
      channel and filament are marked by the arrows. In the first
      column, the contours of line-of-sight magnetic field are shown
      with green for 500~G and yellow for $-500$~G.}}
  \label{f1}
\end{figure*}

\section{Data and Method} \label{1}
\subsection{Event and Data}
AR 12371 owned a complex magnetic field configuration and launched
four successive fast CMEs during its disk transit from 2015 June 18 to
25. These CMEs were associated with long-duration M-class flares on
June 18, 21, 22, and 25, respectively. On June 21, a halo CME left the
Sun when the AR was near the disk center (N12, E16) and generated a
strong geomagnetic storm ($D_{st}$ index was $-204$~nT) on June
22. \textbf{The CME was associated with an M2.6 flare, which started at
  around 1:00~UT on June 21}. A recent study of this AR by
\citet{2018Vemareddy} presented an analysis of the 3D magnetic field
extrapolation with NLFFF model and studied these CMEs in relation to
the coronal magnetic evolution. We pay attention to their dynamic
evolution specifically to disclose the underlying physics of this
complex region. We utilized the EUV imaging data from the Atmospheric
Imaging Assembly (AIA, \citealt{2012Lemen}) onboard the Solar Dynamic
Observatory (\textit{SDO}) to judge the temporal evolution of the AR
at first. The \textit{SDO}/AIA provides full-disk coronal images in 7
EUV filtergrams with \textbf{pixel size of 0.6 arcsec} and a cadence
of 12~s. Observations of the photospheric magnetic field were taken
from \textit{SDO}/Helioseismeic and Magnetic Imager (HMI;
\citealt{2012Schou}). Specifically, we choose the data product of the
Space-weather HMI Active Region Patch (SHARPs, \citealt{2014Bobra}) as
input to our model for driving the evolution of the coronal magnetic
field.

\subsection{Data-driven MHD model}

We employed the data-driven active-region evolution MHD (DARE--MHD)
model~\citep{2016Jiang} to simulate the coronal magnetic field
evolution in response to the evolution of the photospheric
magnetogram. In the DARE--MHD model, we solve the full set of 3D,
time-dependent MHD equations with the magnetic field on the bottom
boundary continuously provided by the vector magnetogram from
\textit{SDO}/HMI. The initial condition consists of an extrapolated
NLFFF data (descried in the next section) and a simple atmospheric
model. The plasma is initialized in a hydrostatic, isothermal state
with $T = 10^{6}$~K in solar corona. To imitate the coronal
low-$\beta$ and high tenuous conditions, the plasma density is
configured to make the plasma $\beta$ less than $0.1$ in most of the
computational volume. At the bottom boundary, the plasma density and
temperature is fixed, while the velocity is also inputted from
observation-derived data using the DAVE4VM
method~\citep{Schuck2008}. We use a non-uniform grid with adaptive
resolution based on the spatial distributions of the magnetic field
and current density, which is designed to save computational resources
without losing numerical accuracy \citep{2017Jiang}. The smallest grid
is $\Delta x = \Delta y = \Delta z = 0.72$~Mm. More details of the MHD
simulation model can be found in \citet{2016Jiang, 2018Jiangb}.

\subsection{NLFFF Extrapolation Model}
For \textbf{a} DARE-MHD simulation, an initial coronal magnetic field
is needed. Here, the coronal magnetic field is extrapolated by the
CESE--NLFFF code developed by \citet{2013Jiangb}. This model is based
on an MHD-relaxation method which seeks \textbf{an} approximately
force-free equilibrium. It solves a set of modified zero-$\beta$ MHD
equations with a friction force using an advanced
conservation-element/solution-element (CESE) space-time scheme on a
non-uniform grid with parallel computing \citep{2010Jiang}. The code
also utilizes adaptive mesh refinement and a multi-grid algorithm to
optimize the relaxation process. This model has been tested by
different benchmarks including a series of analytic force-free
solutions \citep{1990LowLou} and MFR models~\citep{1999Titov}. The
results of extrapolation reproduced from \textit{SDO}/HMI are in good
agreement with corresponding observable features like filaments,
coronal loops, and sigmoids \citep{2013Jianga, 2014Jiang}.

\subsection{Magnetic Field Analysis Tools}
The magnetic field data from our simulation is examined in several
aspects including calculation of magnetic twist number $T_{w}$ for
defining the MFR, decay index $n$ of the strapping field that confines
the MFR, as well as magnetic squashing degree $Q$ which can used to
locate critical thin layers where magnetic reconnection might take
place. The magnetic twist number $T_{w}$ is defined
by~\citep{Berger2006}
\begin{equation}
  T_{w} =  \int_{L} \frac{(\nabla \times \mathbf{B})\cdot \mathbf{B}}{4\pi B^2} \,dl,
\end{equation}
where $L$ is along the magnetic field lines starting from one
footpoint to the other on the bottom boundary. $T_{w}$ measures the
number of winding turns between two infinitesimally close field
lines~\citep{2016LiuR}, and clearly $T_{w}$ is \textbf{a global
  parameter for} any given field line. Here we compute $T_{w}$ for the
whole 3D volume and then the MFR can be identified by coherent group
of field lines with $T_{w} \ge 1$ (or $T_{w} \le -1$). Thus by showing
a isosurface of $|T_{w}|=1$ we can easily find the MFR in the full 3D
volume~\citep[e.g.,][]{DuanA2019}. The decay index $n$ is calculated,
which is defined by
\begin{equation}
  n = - \frac{\partial(\log B)}{\partial(\log h)}.
\end{equation}
It quantifies the spatial decaying speed of the strapping field
strength $B$ with distance $h$ from the bottom surface. Here the
strapping field is approximated by the potential field model
extrapolated from the $B_z$ component of the photospheric magnetogram,
and particularly, only the horizontal component $B_{\rm h}$ of the potential field
is used as being the strapping field $B$. \textbf{It would be more accurate if use only one
component of $B_{\rm h}$ perpendicular to the PIL as the strapping field, because the MFR's axis is roughly parallel to the PIL. But for a potential field, its horizontal field is nearly perpendicular to the PIL. So computing decay index using total $B_{\rm h}$ should be close to that using only the perpendicular component.} According to previous works, the
torus instability of the constrained MFR will be triggered when it
enters a domain with $n \gtrsim 1.5$ \citep{1978Bateman,2006Kliem}. We also
derive the magnetic squashing degree ($Q$ factor) based on the mapping
of two footpoints for a field line \citep{2006Demoulin}. This
parameter can quantify the change of the field line linkage and locate
prominent magnetic separatrix and thin layers, known as
quasi-separatrix layers (QSL), where magnetic field-line mapping
changes abruptly and 3D magnetic reconnection is likely to occur
(e.g. \citealt{1995Priest&Demoulin, 2010Aulanier,
  2006Demoulin}). Additionally, we use the distribution of ratio of
current density to magnetic strength, $J/B$, to locate thin current
layer in the simulation data. It has been shown that $J/B$ is a better
indicator that can highlight current sheet-like distribution than the
$J$ itself \citep{2006Gibson, 2007Fan, 2016Jiang}.

\begin{figure*}
  \centering
  \includegraphics{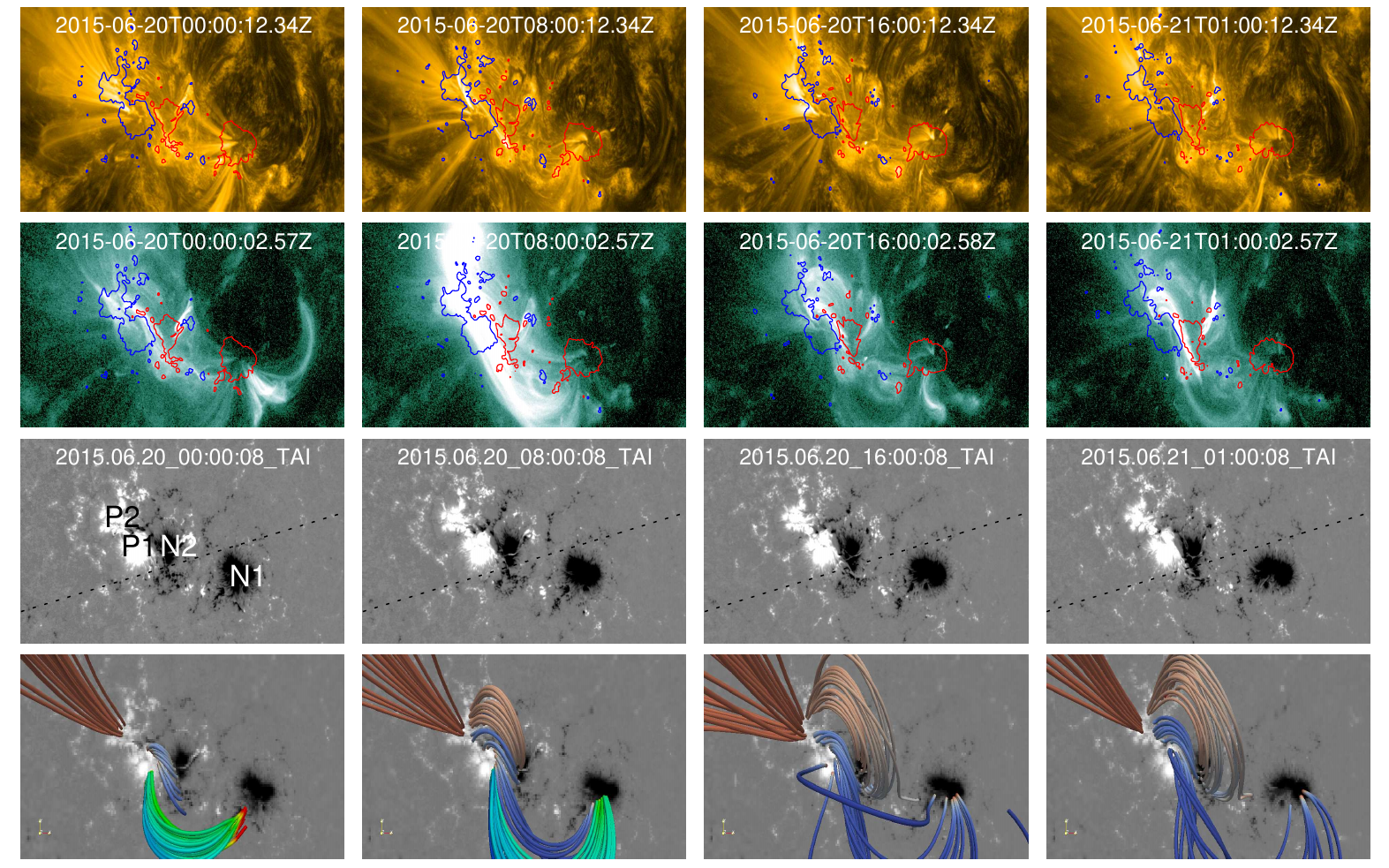}
  \caption{Comparison between the observed coronal loops of the AR and
    the modeled magnetic configuration. The first and second rows show
    EUV observations of AIA 171 \AA{} and 94 \AA{},
    respectively. Contour lines for $B_{z}$ = 500 G (blue) and -500 G
    (red) are overlaid. The third rows show the development of
    photospheric magnetic field observed by SDO/HMI. The last rows
    show sampled 3D magnetic field lines at corresponding time
    overlaid on the background magnetogram. The field of view of AIA
    observations are coaligned with HMI observations.}
  \label{f2} 
\end{figure*}

\section{Results} \label{2}
\subsection{Basic configuration of the pre-flare corona}
Firstly we show the observed erupting structure of the flare. In
Figure~\ref{f1}, simultaneous observations of \textit{SDO}/AIA in
different wavelengths, including 94, 304 and, 131 \AA{}, present the
evolution immediately before and during the eruption.  From these
observations, it can be seen that there is a large-scale, although
rather faint, hot channel erupting toward the southern direction
during the flare. The hot channel, as marked by the arrows in the
figure, connects the east sunspot with the one in the southwest and
forms an inverse J shape with its northeast part slightly hooked. Also
there is an erupting filament of similar shape as shown in the images
of AIA 304 \AA{}. The presence of such hot channel as well as filament
are often deemed to be manifestation of a corresponding MFR
\citep{2012Cheng, 2012Zhang}. More details of observation for this
flare and eruption can be found in \citet{2017Vemareddy} and
\citet{2018Vemareddy}.

Figure~\ref{f2} compares a sequence of the photospheric magnetic
configurations around the flare source region with \textbf{the}
corresponding EUV observations. We also plot sampled magnetic field
lines derived from the DARE--MHD model and attached to the
photosphere. From the photospheric magnetogram, we notice that this AR
consists of a negative unipolar spot in the west and a bipolar spot
group in the vicinity. Here the major magnetic polarities in the
photosphere are marked as P1, P2, N1 and N2. A set of twisted field
lines take shape of sigmoid structure lying along the PIL, which is in
accordance with EUV observations and in turn suggests the existence of
the MFR again, which is analyzed in details below. It's also worth
mentioning that several field lines connect the positive region in the
east to the negative region in the west, which matches well with the
structure of coronal loops in 94 \AA{} images.

\begin{figure*}
  \centering
  \includegraphics[width=0.8\textwidth]{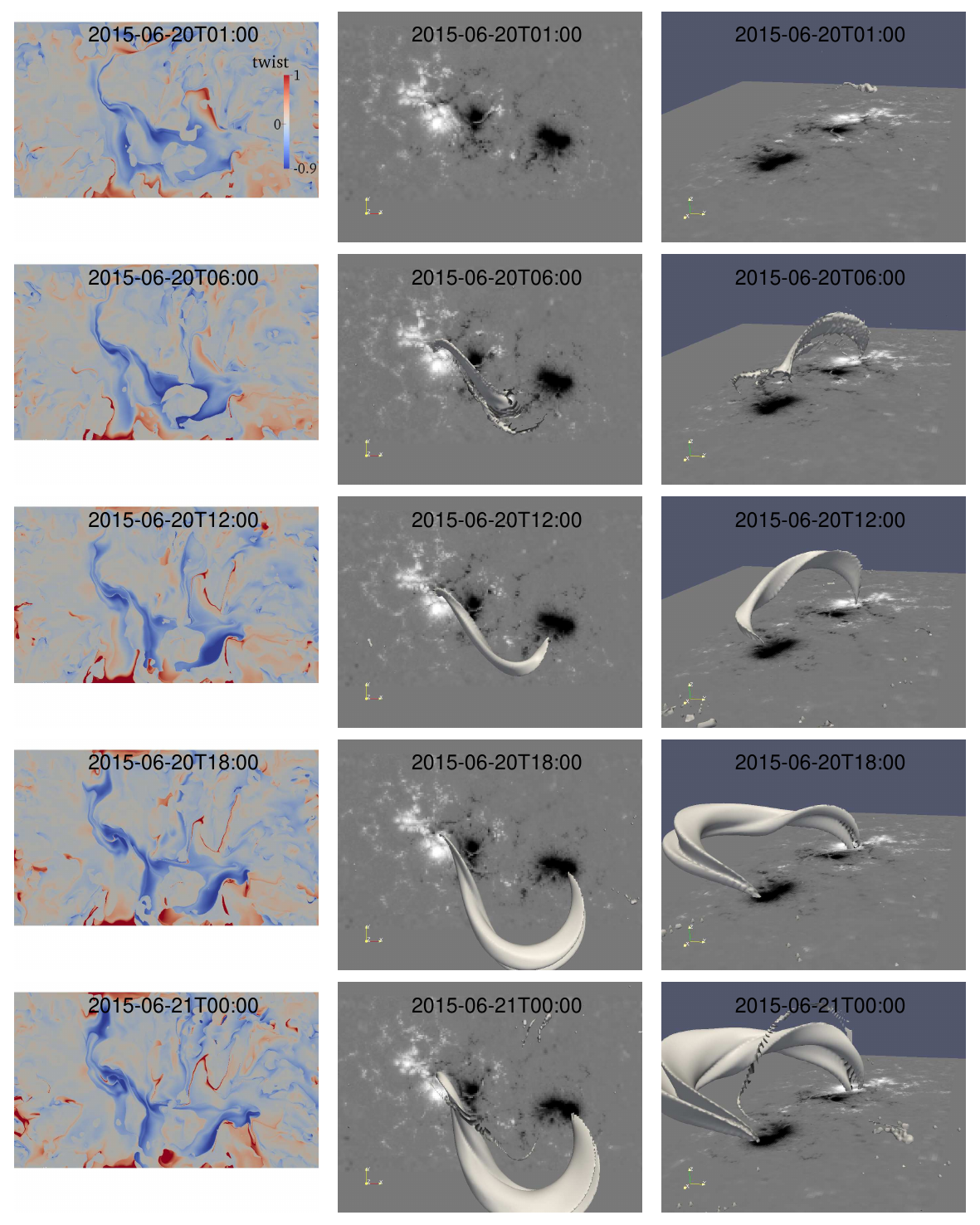}
  \caption{Evolution process of the MFR from a series of simulation
    time and different viewpoints. First column: distribution of
    $T_{w}$ on the photosphere. Second column: \textbf{3D structure of
      the MFR as shown by the isosurface of $T_{w}=-1$} overlaid on
    the background of photospheric $B_{z}$ map. Note that the map of
    magnetic twist number $T_w$ in the first panel has the same field
    of view of the photospheric magnetogram. The third column shows a
    side view of the same structure in the second column.}
  \label{f3}
\end{figure*}

\subsection{Formation of a long flux rope}
The DARE--MHD simulation provides an important insight into the
formation process of the long MFR. The initial time of the simulation
is 00:00~UT on June 20. Time-dependent evolution of the magnetic
structure in five time snapshots is shown from both top view and side
view in Figure~\ref{f3}. Here we use isosurfaces with twist number
$T_{w} = -1$ to show the position of the MFR, which is defined by a
bundle of coherent twisted magnetic field lines with twist number
above one turn. At the beginning, there is no MFR seen above the
photosphere as the twist number $T_{w}$ is generally lower than one,
thus, only sheared arcades around the PIL. Along with the dynamic
evolution of the magnetic configuration, an MFR was gradually
generated in the corona. Our simulation of the long MFR agrees well
with the observation of EUV hot loops at \textbf{01:47} presented in
Figure~\ref{f1}. The magnetic twist number ($T_{w}< 0$) is
significantly enhanced along two sides of the main PIL. As shown in
the second and third column of Figure \ref{f3}, overall the MFR
continues to expand upward and extend its legs to two polarities far
apart (N1 and P1 labelled in Figure \ref{f2}). During the formation
process of the MFR, there is hardly any emerging flux in this region
observed from the magnetogram, and actually the total unsigned flux
decreases in the duration~\citep[e.g., see][]{2017Vemareddy}. So the
long curved MFR can only be built up in the corona driven by the
bottom surface motion rather than direct emergence. We suppose that
the magnetic field lines around the MFR may intertwine and reconnect
with each other through tether-cutting reconnection \citep{1999Titov}
after expansion and turn out to be a long MFR.

\begin{figure*}
  \centering
  \includegraphics[width=0.8\textwidth]{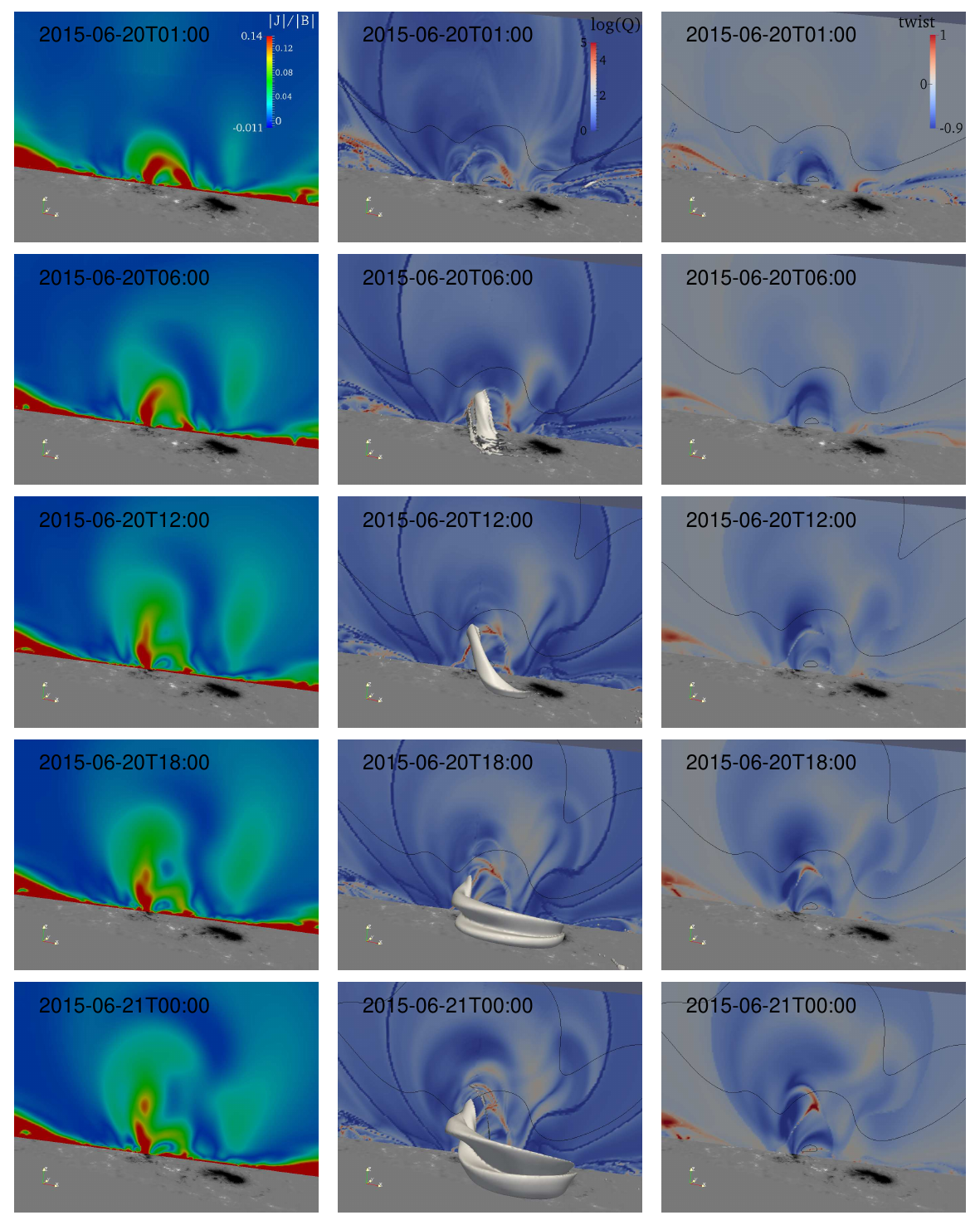}
  \caption{Evolution of magnetic structures in a vertical cross
    section that is along the line marked in Figure~\ref{f2}. All
    panels have the same viewing angle with photospheric $B_{z}$ map
    in the bottom. First column: distribution of the current density
    $J$ (normalized by magnetic field strength $B$) at different
    time. Second column: distribution of magnetic squashing degree
    $Q$. The narrow layers with high $Q$ (in red colors) are locations
    for magnetic topology separatrices and QSLs where magnetic field
    line mapping can change quickly due to reconnection. Third column:
    map of twists number $T_{w}$. The black contour lines of decay
    index $n=1.5$ are plotted in the second and third columns.}
  \label{f4}
\end{figure*}

\begin{figure}
  \centering
  \includegraphics[width=0.45\textwidth]{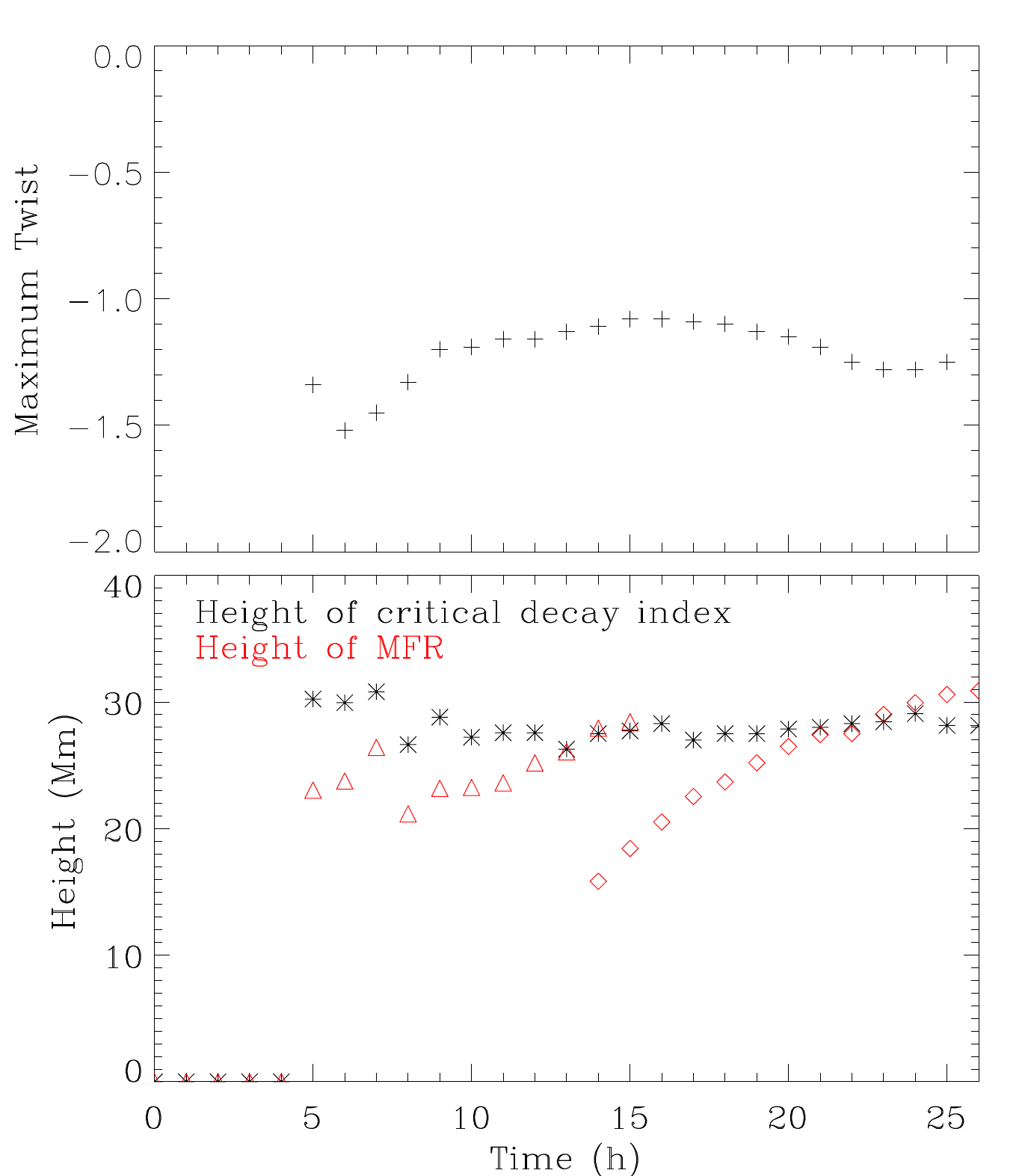}
  \caption{Evolution of the magnetic twist number and the
    height of the MFR as calculated from the vertical slice shown in
    Figure~\ref{f4}. Top panel: the maximum of the twist number in the
    rope, i.e., the twist number of the rope's axis. Bottom panel: the
    height of the apex of the rope axis, and the \textbf{height with
      critical decay index} equals 1.5. Note that at the time of
    $t=13$ the flux rope is split into two parts, and the triangles
    denotes the height of the upper one while the diamonds denotes the
    height for the lower one.}
  \label{f6}
\end{figure}

\begin{figure}
  \centering
  \includegraphics[width=0.45\textwidth]{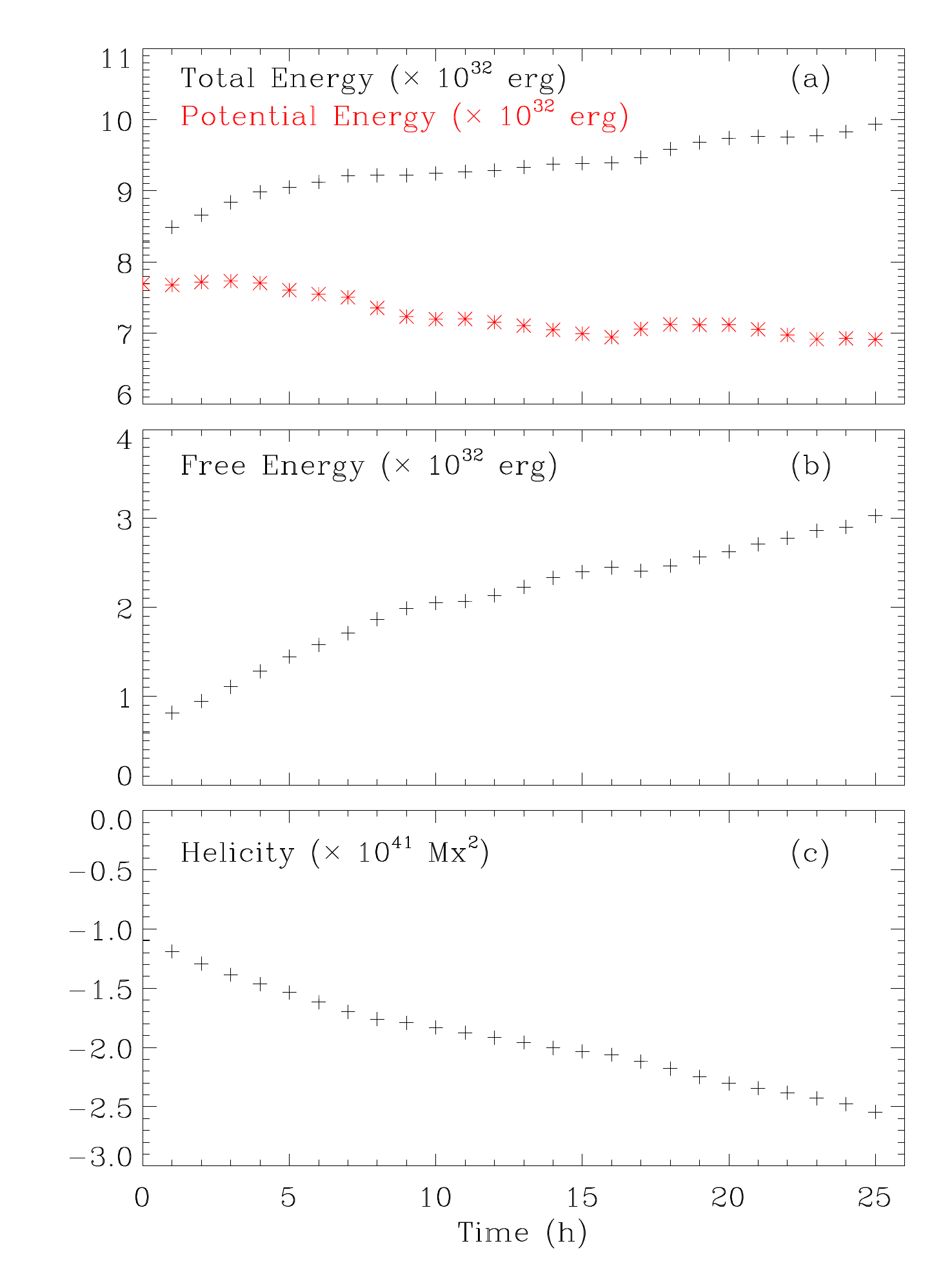}
  \caption{Evolution of magnetic energy (black, derived from
    the MHD model) and potential energy (red, derived from the
    potential field model), magnetic free energy and relative helicity
    from the data-driven MHD simulation.}
  \label{f5}
\end{figure}

Further study on the evolution process of the MFR is revealed in
Figure~\ref{f4} in a vertical cross section which is perpendicular to
the photosphere and along the dash lines as marked in
Figure~\ref{f2}. We also plot the distribution of the $J/B$ (current
density normalized by magnetic field strength) in the first column of
Figure~\ref{f4} for a better analysis of the current density. From the
$Q$-factor maps in the second column, one can see that a QSL (with
high $Q$ value) first forms above the PIL and then the MFR comes into
being. With the generation of MFR, the QSL is further enhanced below
the rope. This clearly suggests that the formation of the rope
resulted from the reconnection in the QSL through a tether-cutting
reconnection in the corona, rather than by a flux cancellation process
where reconnection occurs in the photosphere. Indeed, along the QSL
there forms a thin layer of strong current density, i.e., a current
sheet that is associated with the reconnection, as can be seen in the
first column of Figure \ref{f4}. Matching current concentrations with
the location of main QSLs, an MFR can be separated from its
surroundings \citep{2010Aulanier,2012aSavcheva, 2012bSavcheva}. It is
worth noting that the formation of the MFR can also be clearly seen
from the evolution of the current density. For example, in the last
panel of the first column of Figure \ref{f4}, the sites of strong
currents break into two parts while one part gradually rises up as a
coherent circle, which corresponds to the MFR.

We further calculate the decay index $n$ in the vertical slice. At the
early phase after the MFR formed, it lies relatively low and is far
below the critical \textbf{height with} decay index of $n=1.5$. Then
the MFR's axis is continuously lifted up in the corona. As shown in
Figure~\ref{f6}, the maximum magnetic twist number in the MFR, with
the value approximately at $-1\sim -1.5$, does not \textbf{change}
significantly during the evolution, suggesting that kink instability
cannot be triggered. On the other hand, the height of the rope axis
increases, and near the end of our simulation, for instance, at
$t=24$~h, we notice that the major part of the MFR (the flux with
$T_{w}<-1$) reaches a region with $n>1.5$. According to theoretical
studies \citep{2006Kliem} and results of MHD simulations
\citep{2007Fan, 2010Aulanier}, the torus instability (TI,
\citealt{2006Kliem, 2015Myers}) occurs when the apex of the rope
enters a region where decay index $n$ is larger than a threshold of
$\sim$ 1.5. The TI is a kind of driver of MFR eruption, which is a
result of the loss of equilibrium between the ``hoop force'' of the
rope itself and the ``strapping force'' of the ambient field in
idealized model. So in this event, the MFR has already reached an
unstable region where the TI has a great potential to drive the
eruption. However, a conclusion cannot be drawn directly since the TI
theory is derived from idealized MFR configurations and realistic
coronal magnetic field is much more complex. An interesting fact is
found that the MFR is split into two parts during its evolution. As
can be seen in the bottom panel of Figure~\ref{f6}, the upper part
first reaches the critical height of TI and disappears at time of
$t=14$~h. This splitting of the MFR might trigger a small flare and
eruption. Such phenomena, however, are difficult to analyze in details
with the rather low cadence and low resolution currently used in the
model. There might be a possibility that the splitting \textbf{results
  from} the magnetic island generated in the reconnection, \textbf{as}
a recent observation study shows~\citep{GouT2019}, but still, to
capture correctly the plasmoid in reconnection region requires very
high resolution (such that the aspect ratio of the current sheet can
be very large) which is out of the scope of the current paper.

\subsection{Energies and magnetic helicity evolution}

To further study the global quantities of the magnetic field evolution,
we calculate the magnetic energies and helicity. For instance, the
free magnetic energy ($E_{\rm free}$) refers to the part of magnetic
energy that can be released during eruptions. It can be derived by
subtracting the potential energy ($E_{\rm pot}$) from the total magnetic
energy $E_{\rm tot}$,
\begin{eqnarray*}
  E_{\rm free} & = & E_{\rm total}-E_{\rm pot} \\
           & = & \int_{V} \frac{{B_{\rm tot}}^2}{8\pi} \,dV -
                 \int_{V}\frac{{B_{\rm pot}}^2}{8\pi} \,dV
\end{eqnarray*}
where $V$ denotes the full computational volume of the simulation
(note that here we used the CGS units). The evolution of total energy
and free energy from $t=0$ to $t=26$ (1 hr interval) of the MHD
simulation system are plotted in Figure~\ref{f5}. As can be seen, the
total magnetic energy keeps increasing while the potential field
energy decreases. As a result, the free energy keeps increasing, which
is consistent with the increasing of electric current in the
corona. For the one day evolution, the amount of $E_{\rm free}$ is
accumulated to $\sim 2\times 10^{32}$ erg. Apart from the free energy,
the relative magnetic helicity is also a crucial indicator of the
non-potential nature of the magnetic field \citep{1984Berger},
especially for the existence of MFR.  \textbf{In a closed volume $V$, the
relative magnetic helicity $H$ of a magnetic field $\mathbf{B}$ is defined as \citep{1984Berger, 1985Finn},
\begin{equation}
  H  =  \int_{V}(\mathbf{A}+\mathbf{A}_{\rm p}) \cdot (\mathbf{B}-\mathbf{B}_{\rm p}) d V,
\end{equation}
where $\mathbf{B}_{\rm p}$ is the potential field with same magnetic flux
distribution of $\mathbf{B}$ on the surface of the volume, and $\mathbf{A}$, $\mathbf{A}_{\rm p}$
are corresponding vector potentials of $\mathbf{B}$ and $\mathbf{B}_{\rm p}$, respectively, i.e.,
$\mathbf{B}=\nabla \times \mathbf{A}$, $\mathbf{B_{\rm p}}=\nabla \times \mathbf{A_{\rm p}}$. Here we compute the relative magnetic helicity following the method proposed by \citet{2012Valori}. } As shown in Figure~\ref{f5}(c),
the relative helicity evolves very similar to that of the magnetic
free energy, which is also consistent with the building up and
strengthening of the MFR. The continuous injection of magnetic
helicity and magnetic free energy should be attributed to the driving
of photospheric magnetic field evolution.

\section{Conclusions} \label{3}

Investigation of solar magnetic field configuration and evolution is
essential for the understanding of the nature of solar eruptions. In
this paper, we studied the formation process of an MFR that is
associated with a major eruption event occurring on 2015 June 21 in AR
12371. We performed a data-driven MHD numerical simulation to recreate
the 3D coronal magnetic evolution of this region. The simulation
covers one-day time evolution before the eruption. Our model and
analysis in details combining AIA observation from \textit{SDO} reveal
the formation and evolution process of a long flux rope that is
consistent with EUV observations but is not reconstructed from
previous NLFFF extrapolations. We further computed the magnetic twist
number, decay index, magnetic energy and helicity to investigate how
the MFR originates, is built up and runs into an unstable state that
is likely to \textbf{trigger} its eruption.

Our simulation has successfully and realistically generated the
evolution of an MFR from the vector magnetogram. Compared with the
previous work of \citet{2018Vemareddy} based on the static modeling
through NLFFF method, the result of the self-consistent MHD modelling
offers \textbf{a} unique way to probe the dynamic formation process of
the MFR. We can see the evolution of MFR before the eruption
comprehensively and identify sophisticated structures like QSLs which
may influence MFR's eruption.

Although it is now commonly believed that MFR plays a key role in
solar eruptions, how and when an MFR associated with eruption forms is
still under debates. Here we found that the MFR does not bodily emerge
below the photosphere but forms gradually through magnetic
reconnection in the corona before the flare. Comparison of a sequence
of \textit{SDO}/AIA images with the reconstructed magnetic field
topology confirms the existence of the MFR before the eruption
\textbf{in the analyzed event} and shows the elongation of the
MFR. From the simulation of the pre-flare magnetic topology, it is
found that this long curved MFR was formed gradually above the PIL and
extended out afterwards. Persistent injection of helicity and
accumulation of magnetic free energy provide crucial ingredient for
the building up of the MFR, which may result from the shear and
rotation motion on the photosphere. Calculation of decay index
suggests that the flux rope has reached an unstable region where TI
may trigger the eruption.

In conclusion, all these findings demonstrate the complexity of
pre-flare magnetic topology and \textbf{disclose} the formation and
triggering mechanisms behind a large-scale MFR. This study is
important to understand the role of complex magnetic topology and also
reveal the MFR formation progress before the flare in detail. More
work is supposed to be done on this issue to determine fundamental
triggering mechanisms of the eruption and better characterize the
dynamics of solar eruptions.

\acknowledgments

\textbf{The authors wish to thank the anonymous referee for his
  insightful comments in improving the paper}. This work is supported
by the National Natural Science Foundation of China (NSFC 41822404,
41731067, 41574170, 41531073), the Fundamental Research Funds for the
Central Universities (Grant No.HIT.BRETIV.201901). Data from
observations are courtesy of NASA {SDO}/AIA and the HMI science teams.
The computation work was carried out on TianHe-1 (A) at the National
Supercomputer Center in Tianjin, China. C.W.J. thanks ISSI-BJ for
supporting him to attending the team meeting led by J. C. Vial and
P. F. Chen.




\end{document}